\documentclass[pra,aps,twocolumn,showpacs,epsfig]{revtex4}
\newcommand{\beq}{\begin{equation}}
\newcommand{\eeq}{\end{equation}}
\newcommand{\beqa}{\begin{eqnarray}}
\newcommand{\eeqa}{\end{eqnarray}}
\newcommand{\ba}{\begin{array}}
\newcommand{\ea}{\end{array}}
\usepackage{color}
\usepackage[dvips]{epsfig}
\usepackage{amssymb}
\usepackage{amsmath}

\begin{document}

\title{Entanglement entropy and macroscopic quantum states\\                  
with dipolar bosons in a triple-well potential}

\author{L. Dell'Anna$^1$, G. Mazzarella$^1$, V. Penna$^2$, L. Salasnich$^1$}

\affiliation{$^1$ Dipartimento di Fisica e Astronomia â€Galileo Galileiâ€ and
CNISM, Universit$\grave{a}$ di Padova, Via Marzolo 8, 35122 Padova, Italy\\
$^2$
Dipartimento di Scienza Applicata e Tecnologia and u.d.r. CNISM, Politecnico di Torino,
Corso Duca degli Abruzzi 24, I-10129 Torino, Italy}
\date{\today}

\begin{abstract}
We study interacting dipolar atomic bosons in a triple-well potential within a ring geometry.
This system is shown to be equivalent to a three-site Bose-Hubbard model. We analyze the ground state of dipolar bosons by varying the
effective on-site interaction. This analysis
is performed both numerically and analytically by using suitable coherent-state
representations of the ground state.
The latter exhibits a variety of forms ranging from the su(3) coherent state
in the delocalization regime to a macroscopic cat-like state with fully localized populations,
passing for a coexistence regime where the ground state displays a mixed character.
We characterize the quantum correlations of the ground state from the bi-partition perspective.
We calculate both numerically and analytically (within the previous coherent-state
representation) the single-site entanglement entropy which, among
various interesting properties, exhibits a maximum value in correspondence to the
transition from the cat-like to the coexistence regime.
In the latter case, we show that the ground-state mixed form corresponds, semiclassically,
to an energy exhibiting two almost-degenerate minima.

\end{abstract}


\maketitle

\section{Introduction}
The physics of quantum gases is dramatically affected by interatomic interactions \cite{blochrev}. These, in most cases, are effectively described by a short-range isotropic potential characterized by the s-wave scattering length. However, when the quantum gases are dipolar, then the anisotropic dipole-dipole interaction between magnetic or electric dipole moments becomes very important due to its long-range character and determines novel interesting properties
\cite{baranov}-\cite{muller1}.

Very recently, dipolar bosons trapped by triple-well potentials have became a topic of growing interest. This system often features the open-chain geometry where all condensates are mutually coupled through dipole interaction but only the two lateral condensates exchange bosons with the central one.
Among many interesting aspects, the triple-well system exhibits states revealing the non-local character of the the dipole-dipole interaction within the Josephson-like dynamics \cite{lahaye2} and a complex ground-state phase diagram depending on inter-site and on-site interaction \cite{pfau}.

A different interesting choice for the space structure of the triple well is the ring geometry characterizing the closed chain with periodic boundary conditions. Its realization by means of optical lattices has been designed in Ref. \cite{Amos} and is supported by the recent construction of traps able to confine bosons in toroidal domains \cite{ryu}. Such a geometry has been widely used for investigating the so-called Bose-Hubbard (BH) triple-well where only on-site (nondipolar) interaction are present. The lattice geometry reflects on the dynamics which exhibits a rich variety of translational invariant states and includes ring currents represented by vortex states \cite{vittorio0}-\cite{vittorio1}. The ring geometry, however, gives its most dramatic manifestation in the attractive strong-interaction regime where, owing to the inherent translation symmetry, the ground-state is a Schr\"{o}dinger cat formed by a superposition of macroscopic localized states \cite{vittorio1, vittorio2}. Recently, the transition between
different density-wave phases has been studied for dipolar bosons in a ring-shaped lattice \cite{maik1}.

The ring geometry of the model representing dipolar bosons plays a crucial role in establishing its complete equivalence with the symmetric BH triple-well. Simple calculations show how, with $M=3$ sites, the term representing dipolar interaction can be absorbed by the on-site interaction term.
This process can be effected neither with $M>3$ nor with an open chain. This equivalence gives the  possibility to exploit the considerable amount of information about the dynamics of symmetric triple-well to investigate three-well dipolar models.

In this work, we consider a number $N$ of interacting dipolar bosons at zero temperature confined
by a triple-well potential forming an equilateral triangle. The microscopic dynamics of this system
is effectively described by a 3-site extended Bose-Hubbard (EBH) Hamiltonian which includes the
nearest-neighbor hopping $J$ and the density-density interaction on the same site $U_0$ and between
different sites $U_1$ as well. We diagonalize this Hamiltonian and study the ground-state properties
by varying the effective on-site interaction $U=U_0-U_1$. When this is zero, the ground state of the
EBH Hamiltonian reduces to a su(3) coherent state (describing the boson delocalization) which tends
to a Fock state (with the same boson number in each well) when $U > 0$ becomes large enough.
We show that, for positive $U$, the ground state exhibits the well-known Mott-like (superfluid) form
when the effective interaction is sufficiently large (weak). For negative $U$, the ground state goes
from a su(3) coherent state (where bosons are delocalized) to a macroscopic superposition of three localized states (cat-like state) which, for $|U|$ large enough, evolves towards
the NNN state (superposition of three fully-localized Fock states, representing the generalization of the well-known NOON state).
In the intermediate $U$ range between such states, the ground state is expected to have
a mixed character \cite{vittorio1}. We systematically explore this range determining its extension and
making fully visible the complex form of the ground-state in which a coherent state (describing a uniform
boson distribution) is superimposed to a macroscopic cat-like state.
The  semiclassical analysis of the low-energy scenario in the cat-like regime reveals that the latter
features the coexistence of two minima which can be identified as the classical counterpart of the
ground state with a mixed character.

The variety of ground states previously described can be characterized from a genuine quantum-correlation
point of view. We shall calculate the single-site entanglement entropy $S$, both numerically, by diagonalizing exactly
the EBH, and analytically (in the three ground-state regimes: the NNN state, the coexistence, and the su(3) coherent state)
in the limit of large $N$ by following the same approach of \cite{lucamichele,dellanna}.
In Refs. \cite{dellanna,cats,mazzdell} it was shown that the Schr\"{o}dinger cat state (NOON)
is not the most bi-partite entangled state among the BH Hamiltonian ground-states. Likewise,
we point out that the NNN-like state does not exhibit the maximum bi-partite entanglement entropy:
$S$, actually, attains its maximum value within the coexistence regime. We have analyzed as well the first derivative of $S$ with respect to the interaction by relating its minimum and maximum, respectively, to the emergence of three Fock localized states (around the delocalized state) and the precursor of NNN state.

\section{The model Hamiltonian}

We consider a system of $N$ dipolar dilute interacting bosons trapped by a potential $V_{trap}({\bf r})$
confining bosons in a three-well array. $V_{trap}({\bf r})$ is obtained by superposing
a strong harmonic confinement along axis $z$
with three (planar) potential wells located at the vertices of an equilateral triangle with side
$\ell$ in the $x-y$ plane.
Each of these wells is described by the product of two Gaussians with the same width $w$.
Thus $V_{trap}({\bf r})$ reads
\beqa
V_{trap}({\bf r})= \frac{m}{2}\,\omega_{z}^2 z^2
-V_{0}\, \sum^3_{i=1} \exp\bigg(-\frac{2\,({\vec r} - {\vec r}_i )^2 }{w^2}\bigg)
\eeqa
with $m$ the mass of each boson, $\omega_z$ the trapping frequency in the axial direction, $V_0$
the depth of each well and ${\vec r}_1 = (\ell/2, 0)$, ${\vec r}_2 = (0, {\sqrt 3}\ell/2)$
and ${\vec r}_3 = (-\ell/2, 0)$.
We focus on the case $w \ll \ell$ entailing strong localization in the proximity of sites ${\vec r}_i$.

The boson-boson interaction potential
$V({\bf r}-{\bf r}') = V_{sr}({\bf r}-{\bf r}')+V_{dd}({\bf r}-{\bf r}')
$
is the sum of a short-range contact potential $V_{sr}({\bf r}-{\bf r}')= g\delta({\bf r}-{\bf r}')$
and a long-range dipole-dipole potential
%
%
\beqa
\label{interaction}
V_{dd}({\bf r}-{\bf r}')=\gamma \frac{1-3 \cos^2 \theta}{|{\bf r}-{\bf r}'|^3}\, .
\eeqa
Here $g=4\pi\hbar^2 a_s/m$ with $a_s$ the interatomic s-wave scattering length,
$\gamma=\mu_0\mu^2/4\pi$ for magnetic dipoles ($\mu_0$ is the vacuum magnetic susceptibility and $\mu$ is the magnetic dipole moment) or $\gamma=d^2/4\pi \varepsilon_0$ for electric dipoles ($\varepsilon_0$ is the vacuum dielectric constant and $d$ is the electric dipole moment).
For sufficiently large external (electric or magnetic) fields the boson dipoles are aligned along the same direction,
so that $\theta$ is the angle between the vector ${\bf r}-{\bf r}'$ and the dipole orientation.

The model describing dipolar bosons can be derived from the bosonic-field Hamiltonian
\beqa
\label{secondq}
&&\hat{H}=\int d^{3} {\bf r}\hat{\Psi}^{\dagger}({\bf r})\,H_0\,\hat{\Psi}({\bf r})\nonumber\\
&+&\int d^{3}{\bf r}\,d^{3} {\bf r'}\hat{\Psi}^{\dagger}({\bf r})\hat{\Psi}^{\dagger}({\bf r'})V({\bf r}-{\bf r}')\hat{\Psi}({\bf r'})\hat{\Psi}({\bf r})\nonumber\\
&&H_{0}=-\frac{\hbar^2}{2m}\nabla^2+V_{trap}({\bf r})
\; ,\eeqa
where operators $\hat{\Psi}({\bf r})$ and $\hat{\Psi}^{\dagger}({\bf r})$ annihilates and creates a boson
at point ${\bf r}$, respectively. Operator  $\hat{\Psi}({\bf r})$ can be expanded in terms of space modes
\beq
\label{bosonicfield}
\hat{\Psi}({\bf r})=\phi_1({\bf r})\,\hat{a}_1+\phi_2({\bf r})\,\hat{a}_2+\phi_3({\bf r})\,\hat{a}_3
\; , \eeq
where space modes $\hat{a}_{k}$, $\hat{a}^{\dagger}_{k}$
satisfy the Heisenberg-Weyl algebra $[\hat{a}_{k},\hat{a}^{\dagger}_{q}]=\delta_{kq}$ while,
due to the form of the trapping potential, single-particle wave functions $\phi_{k}({\bf r})$
take the form
\beq
\label{singleparticle}
\phi_{k}({\bf r})=g(z)\,w_k(x,y)
\eeq
in which $g(z)$ represents the ground-state wave function of harmonic potential $(m \omega_{z}^2/2) z^2$, and $w_k(x,y)$ is
a single-particle wave function localized in the $k$th well.
Functions $w_k(x,y)$ satisfy the orthonormality condition
$\int dx dy \, w_{k}^{*}(x,y)\, w_{l}(x,y)=\delta_{kl}$ since they are suitable linear combinations of the
three lowest energy eigenfunctions of the (planar part of) triple-well potential $V_{trap}$.
Hence $\int d^{3}{\bf r}\phi_{k}^{*}({\bf r})\phi_{l}({\bf r})=\delta_{kl}$.

By assuming symmetric wells, the resulting dipolar-boson model is
described by the effective 3-site extended Bose-Hubbard (EBH) Hamiltonian
$$
\hat{H} = -J\big[\hat{a}^{\dagger}_1\hat{a}_2
 +\hat{a}^{\dagger}_2\hat{a}_1+\hat{a}^{\dagger}_2\hat{a}_3
 +\hat{a}^{\dagger}_3\hat{a}_2+\hat{a}^{\dagger}_1\hat{a}_3+
 \hat{a}^{\dagger}_3\hat{a}_1\big]\nonumber\\
$$
\beqa
\label{threemode}
+\frac{U_0}{2} \sum^3_{i=1} \hat{n}_i (\hat{n}_i-1)
+
U_1 \big[\hat{n}_1 \hat{n}_2+\hat{n}_2 \hat{n}_3+\hat{n}_1\hat{n}_3\big] \;.
\eeqa
In equation (\ref{threemode})
$\hat{n}_{k}=\hat{a}^{\dagger}_{k}\hat{a}_{k}$ counts the number of particles in the $k$th well.

Microscopic processes are described in Hamiltonian (\ref{threemode}) by the three macroscopic parameters
$J$, $U_0$ and $U_1$. The hopping amplitude $J$ is given by
\beq
\label{hopping}
J=-\int d^3{\bf r}\, \phi^{*}_{k}({\bf r})\bigg[-\frac{\hbar^2}{2m}\nabla^2+V_{trap}({\bf r})\bigg]\phi_{l}({\bf r})
\;,\eeq
where $k$ and $l$ ($k\neq l$) are two nearest-neighbor sites.

Concerning the on-site interaction $U_0$, the latter embodies the contributions of
both short-range and dipole-dipole interactions (notice that in a single cloud the dipole-dipole
interaction yields a mean-field energy shift which is strongly dependent on the anisotropy
of this interaction \cite{lahaye2}).
Then, $U_0$ is given by
\beqa
\label{onsite}
&&U_0= g \int d^3{\bf r}\, |\phi_k({\bf r})|^4\nonumber\\
&+&\gamma \int d^3{\bf r}\,d^3{\bf r}'\,|\phi_k({\bf r})|^2V_{dd}({\bf r}-{\bf r}')\,|\phi_k({\bf r}')|^2
\;.\eeqa
The third parameter is the nearest-neighbor interaction $U_1$.
In principle there would be an exponentially small contribution of the short-range potential
since the overlap between the wave functions on two adjacent wells is nonzero.
This, however, can be shown to be completely negligible \cite{lahaye2}. Then one finds
\beq
\label{densitydensity}
U_1=\gamma \int d^3{\bf r}\,d^3{\bf r}'\,|\phi_{k}({\bf r})|^2\,V_{dd}({\bf r}-{\bf r}')|\,\phi_{l}({\bf r})^2
\;.
\eeq
\begin{figure}
\centering
\resizebox{\columnwidth}{!}{
\begin{tabular}{cc}
\epsfig{file=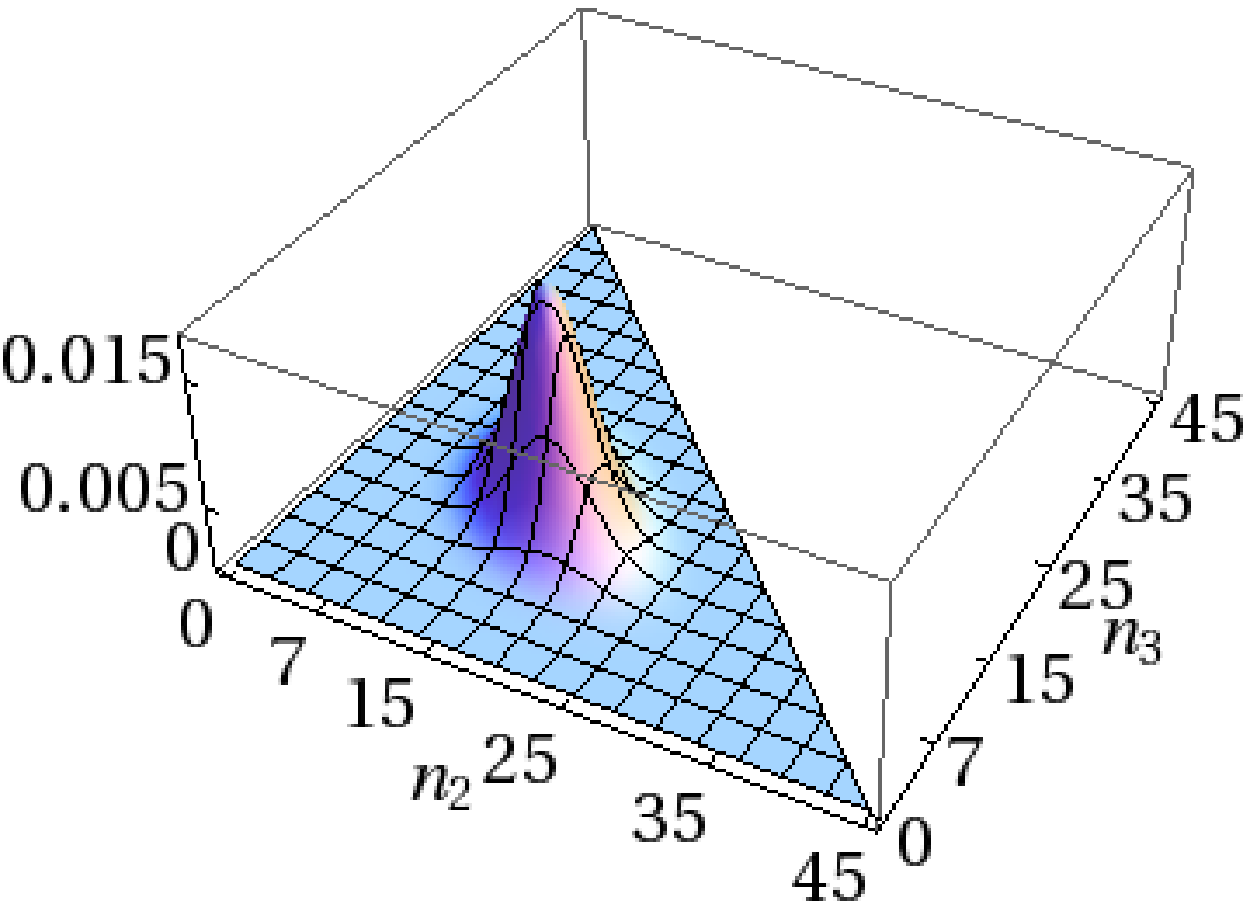,clip=}&
\epsfig{file=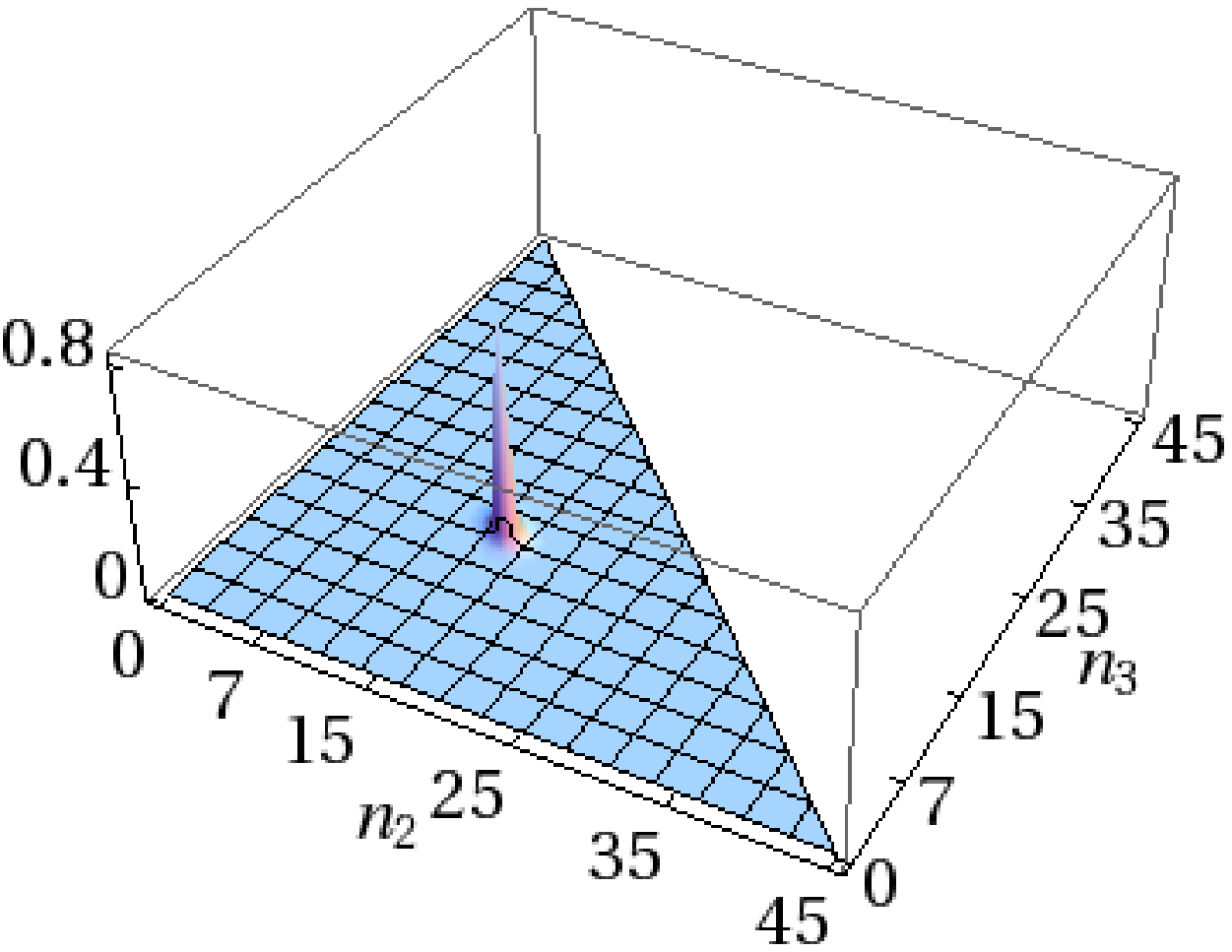,clip=}\\
\epsfig{file=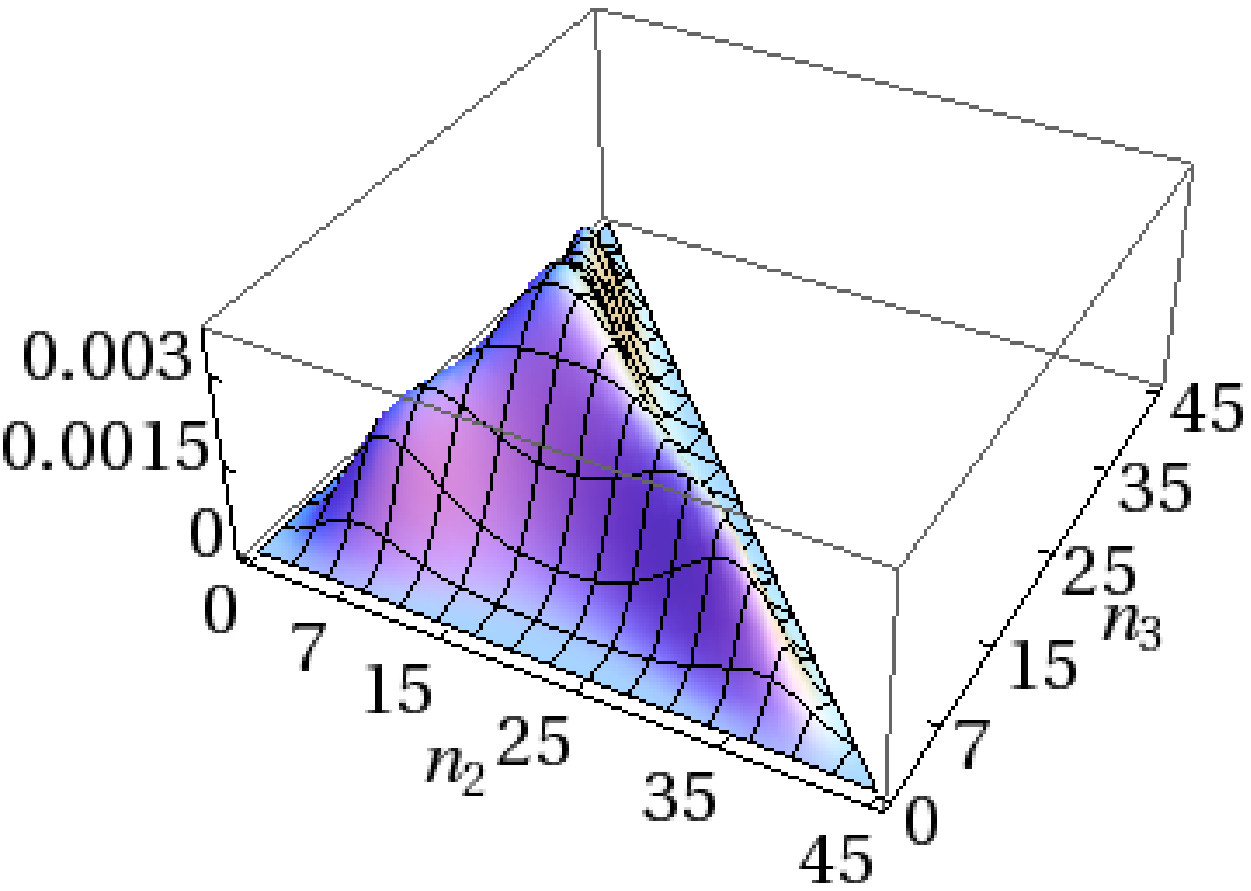,clip=}&
\epsfig{file=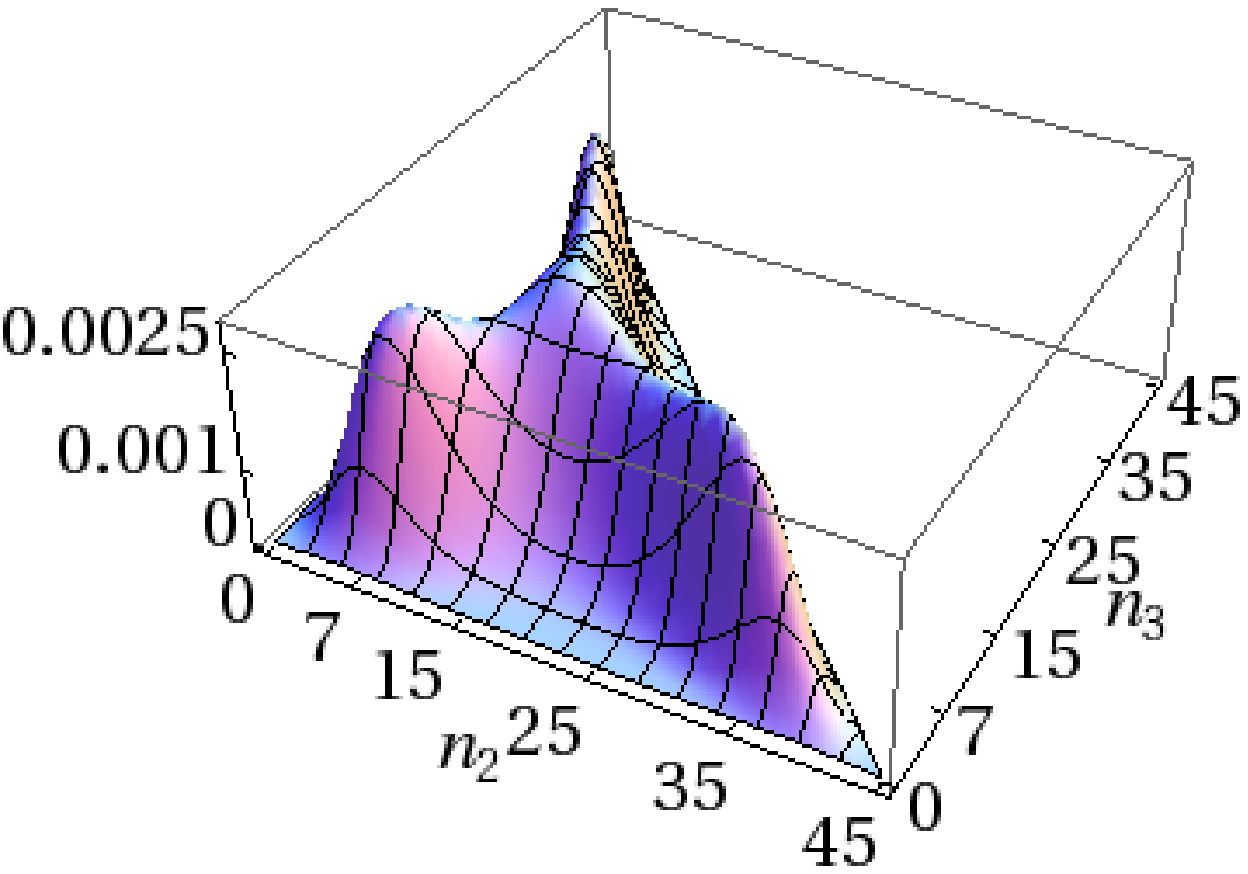,clip=}\\
\epsfig{file=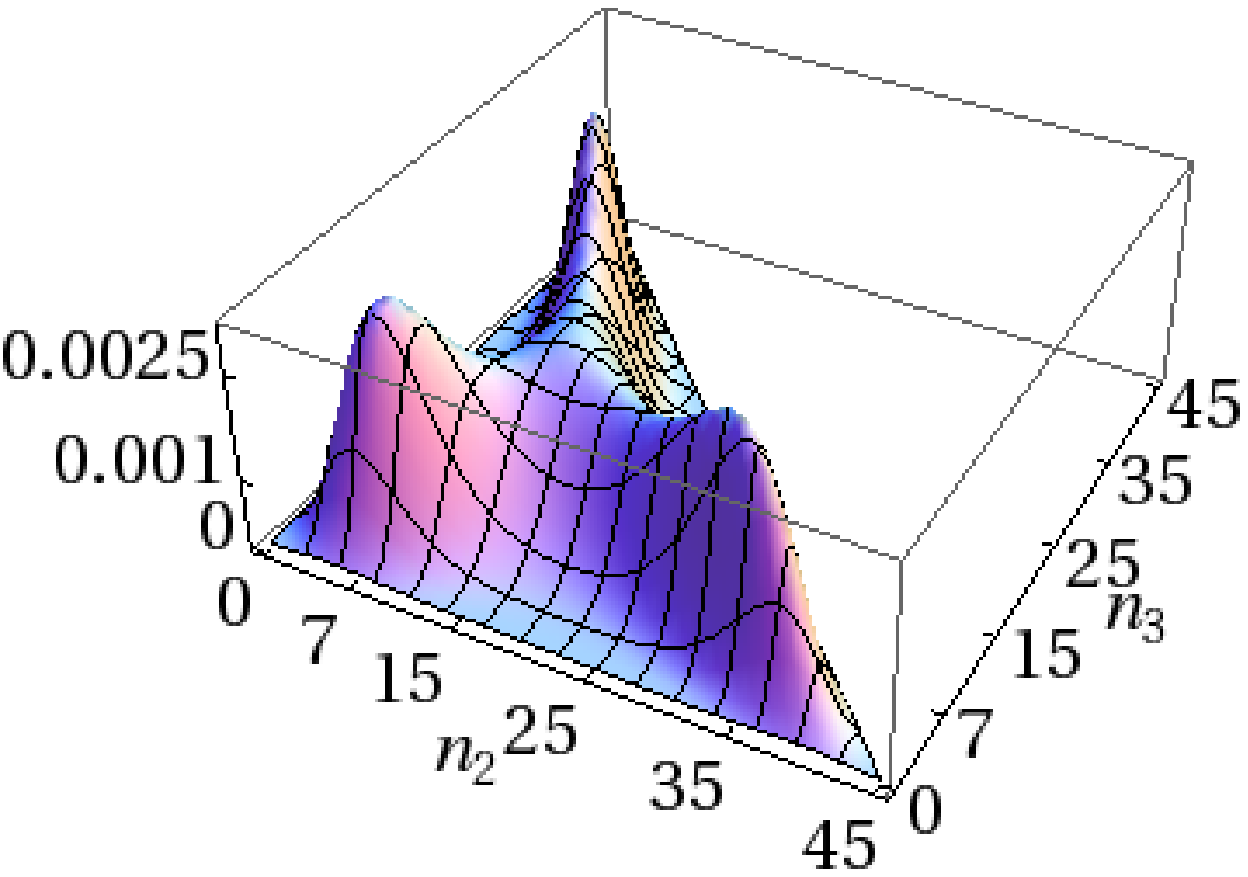,clip=}&
\epsfig{file=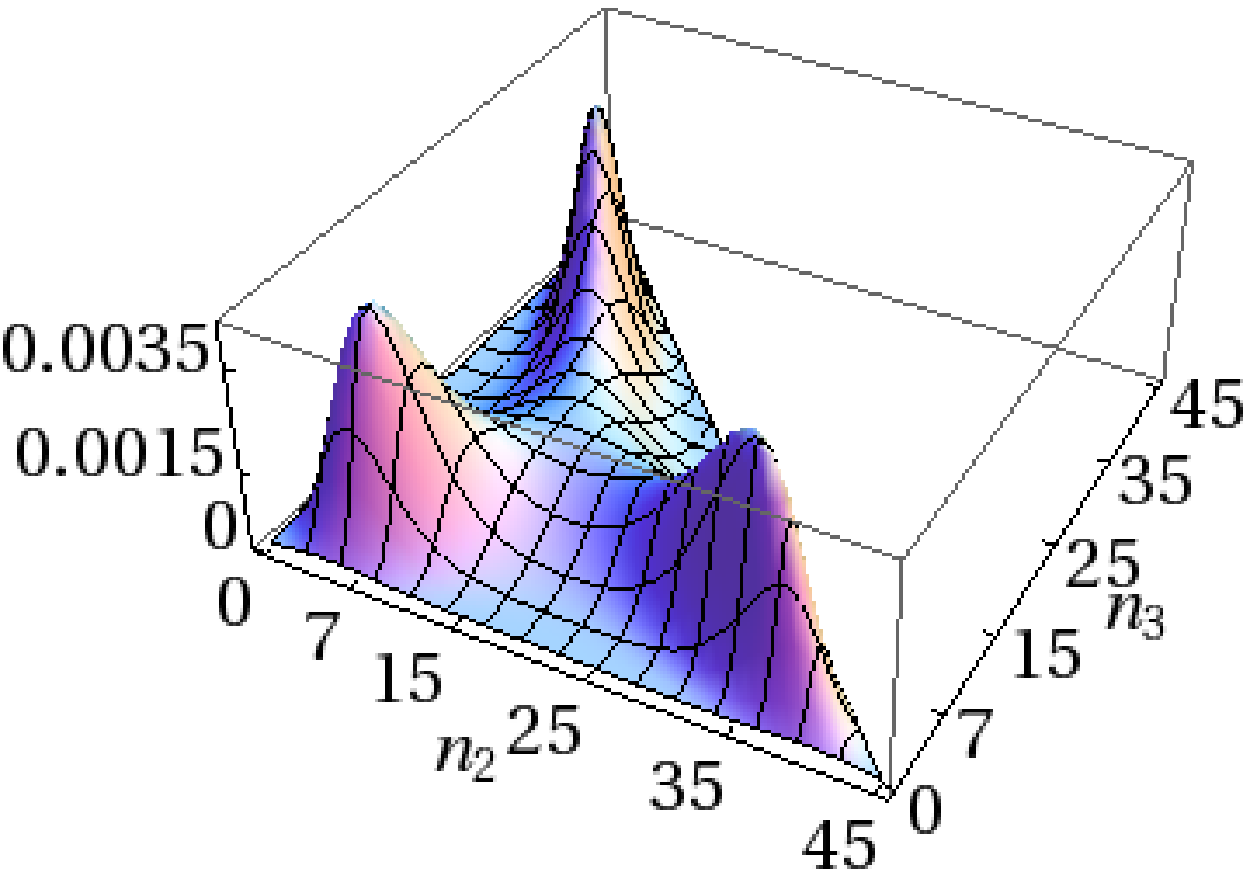,clip=}\\
\epsfig{file=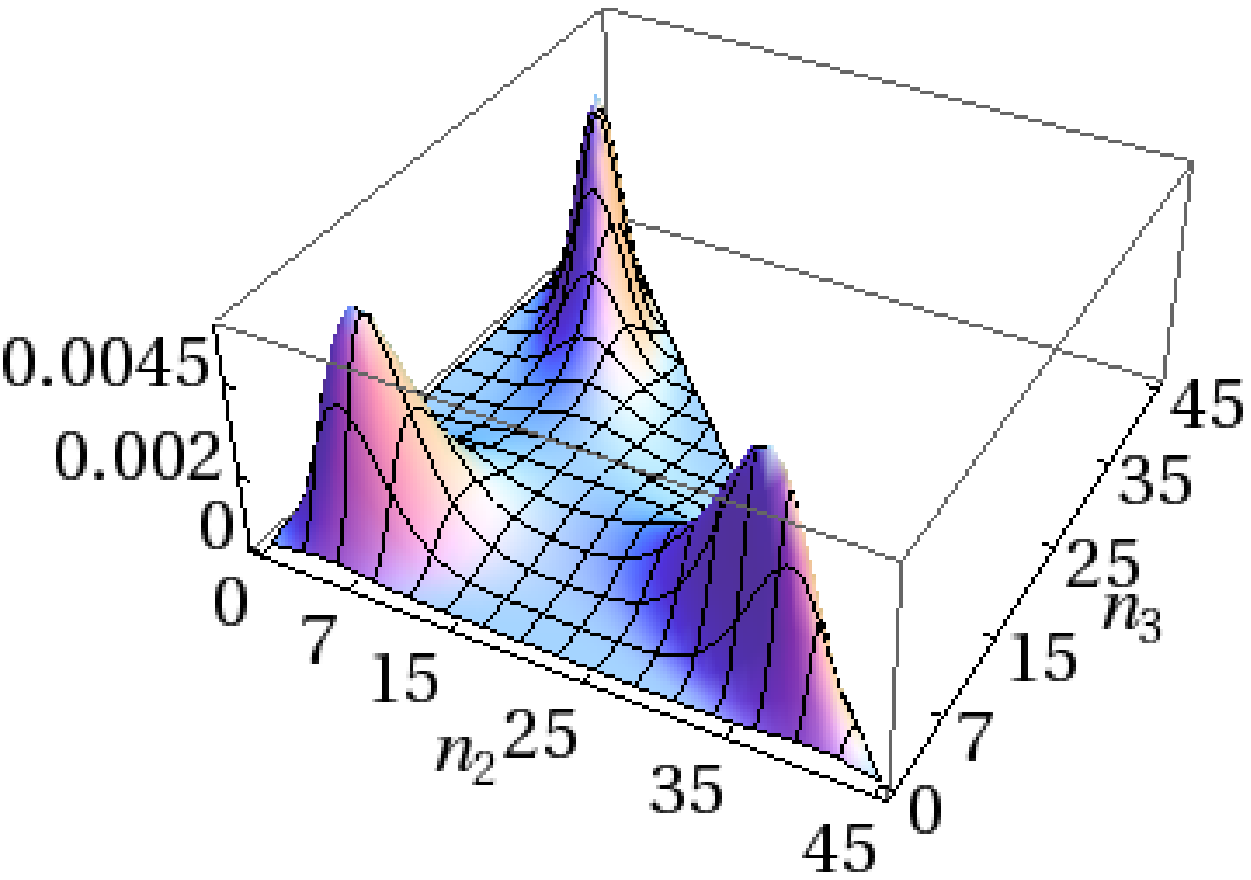,clip=}&
\epsfig{file=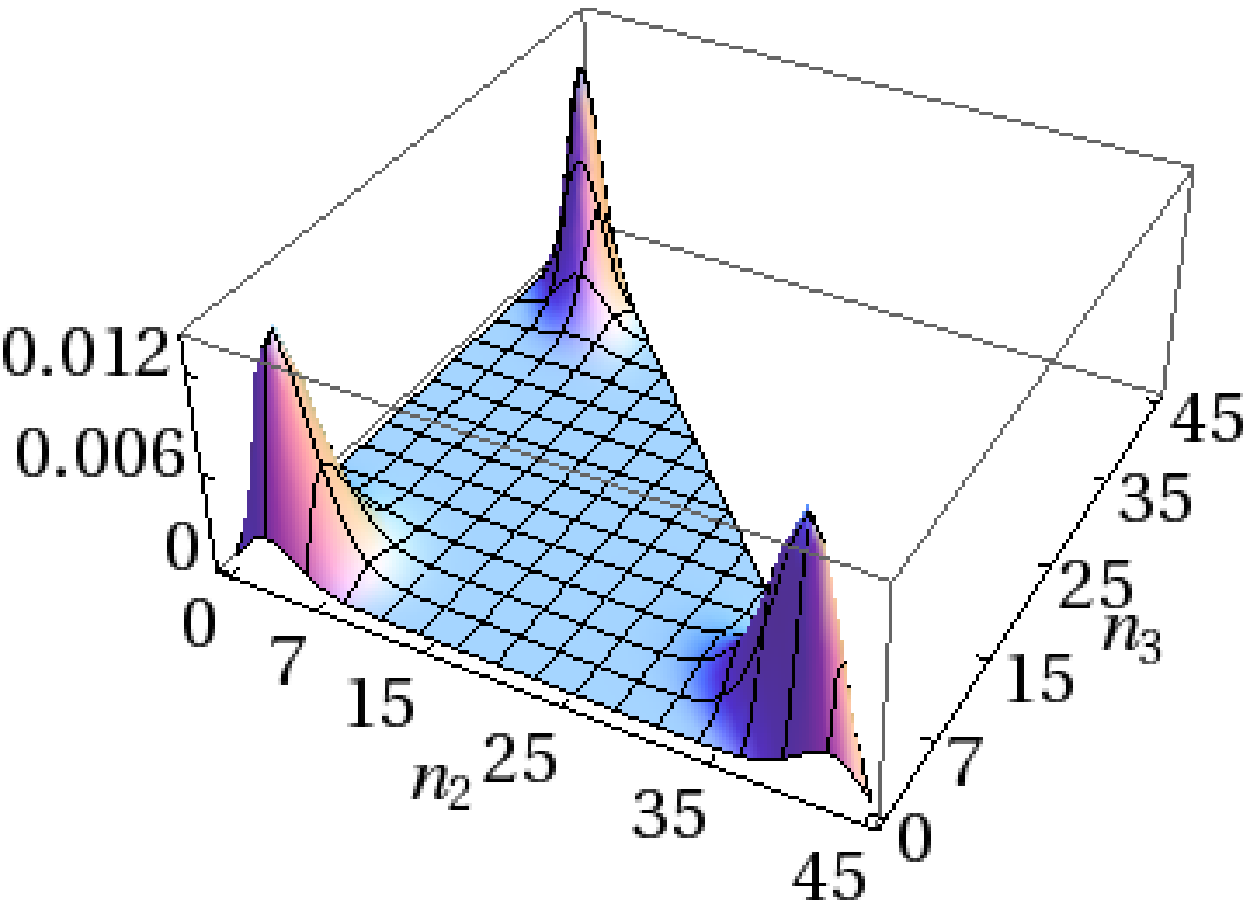,clip=}\
\end{tabular}}
\caption{
\label{fig1} (Color online).
Coefficients $|c_{n_2,n_3}|^2$ as functions of occupation numbers $n_2$ and $n_3$
for different values of $\tau=J/(N|U|)$ and $N=45$ bosons.
Top-bottom (each row, from left to right).
First row (repulsive case $U>0$): $\tau \gg 1$, $\tau=0.000222$.
Second row: $\tau=0.2415$, $\tau=0.2389$. Third row: $\tau=0.2374$, $\tau=0.2349$.
Fourth row: $\tau=0.2315$, $\tau=$ $0.2020$.
The last three rows concern the attractive case $U <0$.}
\end{figure}

Interestingly, owing to the underlying ring geometry, the dipolar-boson model (\ref{threemode})
can be mapped onto a simpler model in which dipolar-interaction terms $\hat{n}_i \hat{n}_k$ are completely removed.
Since
$$
\hat{N}^2=\sum_{i=1}^{3}\hat{n}_{i}^{2}+2(\hat{n}_1\hat{n}_2+\hat{n}_2\hat{n}_3+\hat{n}_1\hat{n}_3)\, ,
$$
where $\hat{N}=\hat{n}_1+\hat{n}_2+\hat{n}_3$ commutes with Hamiltonian (\ref{threemode}),
then $\hat{H}$ reduces to a BH model
$$
\hat{H} = -J\big[\hat{a}^{\dagger}_1\hat{a}_2
+\hat{a}^{\dagger}_2\hat{a}_1+\hat{a}^{\dagger}_2\hat{a}_3
+\hat{a}^{\dagger}_3\hat{a}_2+\hat{a}^{\dagger}_1\hat{a}_3+
\hat{a}^{\dagger}_3\hat{a}_1\big]\nonumber\\
$$
\beqa
\label{effectiveh}
+ \frac{U}{2}\sum^3_{i=1} \hat{n}_i (\hat{n}_i-1)
+\, \frac{U_1}{2} \hat{N}(\hat{N}-1)\;,
\eeqa
where $U \equiv U_0-U_1$ shows that the nearest-neighbor (dipolar) interactions
have been absorbed by the on-site interaction term. In the following, we will neglect the constant
term $U_1\hat{N}(\hat{N}-1)/2$.
Then, the boson-boson interaction can be described in terms of an effective on-site interaction
of amplitude $U_0-U_1$. The latter is repulsive when the $U_0>U_1$ and attractive in the opposite
case. This means (see the discussion below) that making $U_0$ larger or smaller than $U_1$ allows one to explore
a variety of ground-states ranging from the Mott-like state (for $U \gg J/N$ and integer filling)
and attractive/repulsive superfluid states ($|U| < J/N$), involving the same average population per site,
to Schr\"odinger-cat states ($|U| \gg J/N$ with $U<0$) consisting essentially of the superposition
of three fully-localized macroscopic states.

\section{Dipolar-boson ground state}

The ground-state of the Hamiltonian (\ref{effectiveh})
is a superposition of different Fock states which,
due to the conservation of total boson number $N$, can be written as
\beq
\label{superposition}
|\Psi\rangle=
\sum_{n_2=0}^{N}\,\sum_{n_3=0}^{N-n_2}c_{n_2,n_3}|N-(n_{2}+n_{3}),n_{2},n_{3}\rangle
\; .
\eeq
State $|\Psi\rangle$ features different regimes that depend on interaction
parameters $U_0$ and $U_1$ through the effective hopping amplitute $\tau=J/(N|U|)$.

For $\tau \gg 1$, namely if $|U| \ll J/N$ (note that $U_0$ can approach $U_1$ either
from above or from below), the ground state tends to the su(3) coherent state \cite{vittorio3}
\beq
\label{coherentclosed}
|{\rm del} \rangle \equiv |z \rangle = {1\over \sqrt{N!}}
\left ( {\sum}^3_{i=1 } z_i {\hat a}_{i}^{\dagger}
\right )^N |0,0,0\rangle
\; ,
\eeq
where $z_i = 1/\sqrt 3$, and $|0,0,0\rangle$ is the vacuum state
with zero bosons.
State $|{\rm del} \rangle$ describes the delocalization of boson population since the boson-number
expectation value $\langle z |{\hat n}_i |z \rangle = N |z_i|^2 = N/3$ is the same at each well.
Note that in the limit $|U| \to 0$, state (\ref{coherentclosed}) represents an exact eigenstate
of (\ref{effectiveh}). In this case, in fact, Hamiltonian $\hat{H}$ can be rewritten as
$\hat{H}=-J(3\hat{A}^{\dagger}\hat{A}-\hat{N})$
with
$\hat{A}=({\hat a}_{1}^{\dagger} + {\hat a}_{2}^{\dagger}+ {\hat a}_{3}^{\dagger})/\sqrt{3}$
which, owing to its harmonic-oscillator form, is diagonalized by the state (\ref{coherentclosed}).

When $\tau \ll 1$ and $U<0$ the ground state of Hamiltonian (\ref{effectiveh})
tends to a linear combination of the Fock states each one describing the complete localization
of bosons at a single well. Then $|\Psi \rangle \simeq |W\rangle$, where
\beq
\label{w}
|W\rangle=\frac{1}{\sqrt{3}}\big(|N,0,0 \rangle+|0,N,0\rangle+|0,0,N\rangle \big) \;.
\eeq
State $|W\rangle$, hereafter called NNN state, can be interpreted as the $N$-boson
generalization of the three-qubit state \cite{durzeilinger}
$|W \rangle= (|1,0,0 \rangle + |0,1,0\rangle+  |0,0,1\rangle )/{\sqrt{3}}$.
Since each component of $|W\rangle$ features a macroscopic boson localization, such
a state represents a Schr\"odinger cat.

Finally, in the presence of integer filling, for repulsive interaction $U>0$ and $\tau \ll 1$,
ground state (\ref{superposition}) takes the form
\beq
\label{fock}
|\Psi \rangle \simeq |n, n, n \rangle\, , \quad n= \frac{N}{3} \in {\mathbb {N} }_0
\; ,
\eeq
where the majority component $|n, n, n \rangle$
reflects the boson space distribution characterizing the Mott phase.
For $\tau \equiv 0$ one has $|\Psi\rangle \equiv |n, n, n\rangle$. If the filling is non integer
the ground state is well represented by the symmetric form
$|\Psi \rangle \simeq ( |n+s, n, n\rangle +|n, n+s, n\rangle +|n, n, n+s\rangle) /\sqrt 3$.
This describes the expected superfluid character through the presence of a
fully-delocalized particle (hole) excitation when $s=+1$ ($s=-1$).

Fig.~\ref{fig1} shows the squared modulus $|c_{n_2,n_3}|^2$ of the coefficients of $|\Psi\rangle$ components
as a function of occupation numbers $n_2$, and $n_3$. The total number of bosons is $N=45$ entailing integer filling.
The left top panel of Fig.~\ref{fig1}, where $U=0$ describes an infinitely large $\tau $,
represents the su(3) coherent state (\ref{coherentclosed}).
Starting from this configuration, the smaller $\tau$ the narrower the distribution of
$|c_{n_2,n_3}|^2$.
In general, for a generic $\tau \in [0, \infty]$ in the repulsive regime $U>0$,
the quantity $|c_{n_2,n_3}|^2$ attains its maximum value for $n_2=n_3 = 15$.
This situation is evidenced in the right top panel for $\tau=0.00022$, where the Mott-like
ground state is very close to state (\ref{fock}).

For attractive interactions $U<0$ and $\tau \gg 1$ the ground state is,
in general, a delocalized state qualitatively similar to that of
case $U \ge 0 $ with a finite $\tau \gg 1$ (see the left top panel of Fig.~\ref{fig1}).
Below the critical value $\tau_0$ (one has $\tau_0 = 0.25$
for $N$ large enough) ground state (\ref{superposition}) totally modifies its
structure exhibiting a transition \cite{vittorio1}, \cite{vittorio2} to a mixed state
characterized by the coexistence of three symmetric localized states and
a delocalized state (well represented by state (\ref{coherentclosed})).

When $\tau$ is further decreased (top-bottom, second-fourth rows of Fig.~\ref{fig1}),
ground state (\ref{superposition}) once more changes its structure tending to
a Schr\"odinger-cat state essentially coinciding with state (\ref{w}).
Translated into the language of $|c_{n_2,n_3}|^2$, the left panel of
the second row ($\tau=0.2415$) well illustrates the emergence of three
localized states: the distribution of $|c_{n_2,n_3}|^2$, in fact, starts to show
three lateral peaks in addition to the central one describing a uniform
boson distribution.
By further decreasing interaction $U<0$ (see Fig.~\ref{fig1}, second row, $\tau= 0.2389$)
the emergence of three external peaks becomes more and more evident.

The third row of Fig.~\ref{fig1} clearly shows how the stronger the interatomic attraction $|U|$,
the smaller the central peak. In our sequence, the latter reaches its smallest nonzero
value for $\tau=0.2349$ (third row, right panel) and then disappears for $\tau=0.2315$
(fourth row, left panel) entailing a ground state characterized by three well separated
peaks. When $\tau$ becomes small enough, the ground state evolves towards a form well
represented by the NNN state (see the right bottom panel where $\tau=0.2020$).

Some considerations are now in order concerning effective interaction $U = U_0-U_1$.
First, it is worth noting that
for any value of $U$ the same ground state (and more in general any eigenstate
of $\hat H$) can be realized by (in principle) infinitely-many choices of $U_0$
and $U_1$ if such parameters are changed by the same amount.
Interestingly, while in the BH model (where $U_1=0$) the rich structure of the
attractive ground state is obtained if interaction $U_0$ is genuinely negative,
in the EBH model, due to dipolar interactions, the condition $U<0$ can be met with
both $U_0$ and $U_1$ positive. A nontrivial consequence is that one can investigate
the attractive regimes of triple-well dynamics avoiding the possible collapse of the bosonic
cloud expected, for sufficiently large atomic densities, in the presence of truly
attractive bosons.

\section{Entanglement properties of $| \Psi \rangle $}

%
The dramatic changes characterizing the structure of the dipolar-boson ground state when
$\tau$ is varied suggest that quantum correlations between different sites may play an
important role.
To this end we analyze the entanglement properties pertaining to state (\ref{superposition})
for different choices of interaction parameters.
The density matrix $\hat{\rho}$ associated to $|\Psi\rangle$ is
\beq
\label{dm}
\hat{\rho } =|\Psi \rangle\langle \Psi |
\;.\eeq
We describe the quantum correlations by calculating the entanglement between
the well $1$ and the system formed by the wells $2$ and $3$. An excellent measure
of this entanglement is provided by the single-site entanglement (SSE) entropy $S$
\cite{bwae}.
This quantity is the von Neumann entropy of reduced density matrix ${\hat \rho}_{1}$ defined by
\beq
\hat{\rho}_{1} =Tr_{2}Tr_{3} {\hat \rho}
\;.\eeq
The latter is a matrix obtained by partially tracing the total density matrix (\ref{dm}),
first over the degrees of freedom of the well $3$ and then on those of the well $2$.
Therefore, the SSE entropy $S$ is given by
\beq
\label{ee}
S= - Tr_{1} ({\hat \rho}_{1} \log_{2} {\hat \rho}_{1} )
\; .
\eeq

\begin{figure}[ht]
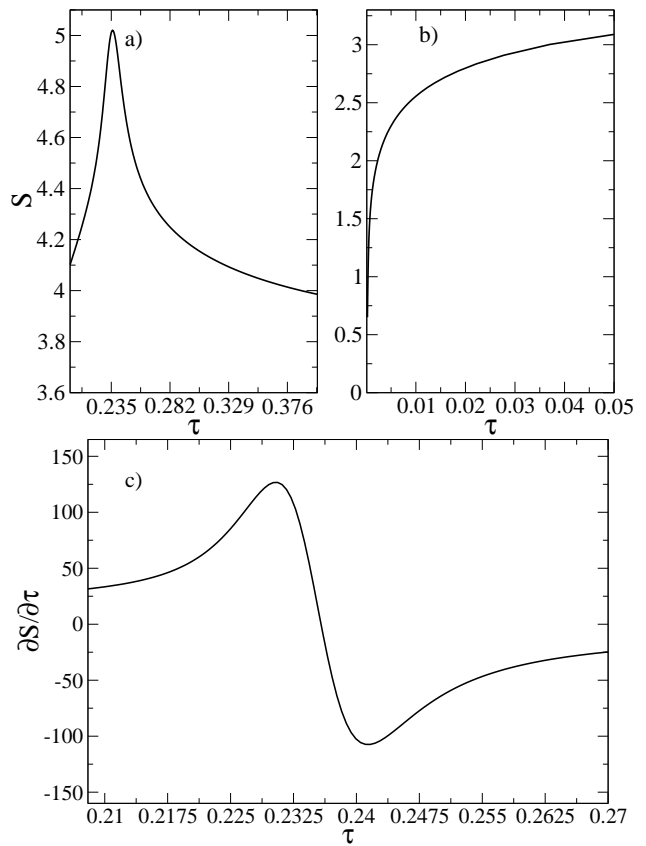

\epsfig{file=ent455.eps,width=0.96\linewidth,clip=}
\epsfig{file=dp5.eps,width=0.94\linewidth,clip=}
\caption{
\label{fig2} Top panels.
The entanglement entropy $S$ (\ref{ee}) vs. $\tau=J/(N|U|)$ for $N$=45 bosons.
Panel a): attractive interaction $U <0$. Panel b): repulsive interaction $U >0$.
Panels c): the first derivative of $S$ vs $\tau$ in the attractive case $U <0$.}
\end{figure}

We have studied
$S$ as a function of $\tau$ both for attractive (Fig.~\ref{fig2}, panel a) and for repulsive
(Fig.~\ref{fig2}, panel b) interactions. For repulsive bosons (Fig.~\ref{fig2}, panel b), $S$ increases when
$\tau$ is increased. $S$ tends to an asymptotic value (not shown) confirming that the ground
state tends, in parallel, for $\tau \gg 1$ to  state $|{\rm del} \rangle$ with a nonzero, constant
quantum-correlation degree reflecting the nonseparable, Gaussian form of $|\Psi \rangle$ in the superfluid
regime. For $\tau \to 0$ the vanishing of $S$ reflects the disentangled form
of Mott-like state $|n,n,n \rangle$ described by a single integer-filling Fock state.

For attractive bosons, $S$ exhibits three remarkably points showed in Fig.~\ref{fig2}, panel c.
For $\tau = 0.2416$, the (first) $\tau$ derivative of $S$ has a minimum. This value of $\tau$ is
very close to that for which the three external peaks crop up (Fig.~\ref{fig1}, second row, left panel)
around the central one inherited from the regime with $\tau \gg 1$.
When $\tau = 0.2360$, the single-site entanglement entropy attains its maximum value ($S=5.0201$).
In correspondence to this value the central peak of the distribution of $|c_{n_2, n_3}|^2$
reaches its smallest nonzero value (Fig.~\ref{fig1}, third row, right panel).
Finally, the first derivative of $S$ has a maximum for $\tau= 0.2306$. This value is very close
to the value of $\tau$ for which the NNN state precursor emerges, that is the state where the
central peak disappears (see Fig.~\ref{fig1}, fourth row, left panel).
Then, the NNN state is achieved by further lowering $\tau$.

It is possible to provide analytical estimations of the SSE entropy in the limit of large $N$.
To this end we consider the generic su(3) coherent state
\beq
|\Psi\rangle
= \frac{1}{\sqrt{N!}}(\xi_1 {\hat a}_1^\dagger+\xi_2 {\hat a}_2^\dagger
+\xi_3 {\hat a}_3^\dagger)^N|0\rangle
\label{state}
\;,\eeq
where $\Sigma_{i=1}^{3}|\xi_i|^2=1$ ensures the normalization condition
$\langle \Psi | \Psi \rangle = 1$.
Within the present coherent-state picture the expectation value
$N_i = \langle \Psi | {\hat n}_i |\Psi \rangle = N |\xi_i|^2 $ of ${\hat n}_i$
can be interpreted as the $i$th-site population.
By defining the relative phases $\phi_{ij}=\theta_i-\theta_j$
\beq
\label{phase}
e^{i \phi_{ij}}=\frac{\xi_{i}\,\xi_{j}^*}{|\xi_i|\,|\xi_j|}
\; , \eeq
the expectation value $E= \langle\Psi|\hat{H}|\Psi\rangle$ of Hamiltonian (\ref{effectiveh})
on the state (\ref{state}) is found to be \cite{vittorio3}
\beq
E= \frac{U}{2} N(N-1) \sum_i |\xi_i|^4 -J N \sum_{k < j} (\xi_k \xi_j^*+ \xi_j \xi_k^*)
\label{scenergy}
\eeq
whose final expression in terms of variables $N_k$ and $\phi_{ij}$ is the
well-known boson Josephson-junction form
$E= \frac{U}{2} \sum_i N_i^2 -2 J \sum_{k < j} \sqrt{ N_k N_j } \cos (\phi_{jk})$
if $(N-1)/N \simeq 1$ is assumed.
Based on ansatz (\ref{state}) we evaluate the SSE entropy (\ref{ee}) of
the ground state for three significant cases: the (uniform) ground state described by state (\ref{coherentclosed}),
the cat-like ground state (including the case of the NNN state (\ref{w})), the regime with the coexistence of different ground states.

\section{analytic estimation of entropy $\it S$ in different
ground-state regimes}

To begin with, we derive the analytic
expression of the reduced density matrix $\hat\rho_\ell$ when $|\Psi \rangle$
is a coherent state (\ref{state}). This expression is obtained by partitioning
the sites in two blocks $[1,\ell]$ and $[\ell+1,3]$, with $\ell=1,2$,
and tracing out the sites belonging to the second block \cite{lucamichele, dellanna}
\beq
{\hat \rho}_\ell=
\sum_{n_{\ell+1},... n_{3}}\frac{1}{\prod_{i=\ell+1}^{3}n_i!} |D (n_{\ell+1},... n_{3}) |^2 \, ,
\eeq
with the constraint $\sum^3_{i=1}n_i=N$ and where
$$
D (n_{\ell+1},... n_{3}) = \langle 0| \prod_{i=\ell+1}^{3}({\hat a}_i)^{n_i}|\Psi \rangle\, .
$$
The resulting $\hat\rho_\ell$ is a $m \times m$ diagonal matrix
with $m=(N+\ell)!/(\ell! N!)$. In particular, for $\ell=1$, the diagonal elements of
this matrix are found to be \cite{lucamichele}
\beqa
\label{densitymatrix}
{\rho}_{1,n} (\xi_1)=\left(
\ba{c}
N\\n
\ea
\right)
\left(1-|\xi_1|^2\right)^{\left(N-n\right)}|\xi_1|^{2n}.
\;\eeqa
This expression gives the probability to find $n$ bosons in the well $1$, and satisfies
the normalization condition $\sum_{n=0}^N {\rho}_{1,n}=1$. In the large-$N$ limit this binomial distribution
can be approximated to a Gaussian one with mean value $N|\xi_1|^2$ and variance
$N|\xi_1|^2|(1-|\xi_1|^2)$. For $N \to \infty $, therefore, the entropy can be calculated
analytically. In this limit, it exhibits the following asymptotic expression \cite{lucamichele}
\beq
S \simeq \frac{1}{2}\log_2 \Bigl ( 2\pi e N|\xi_1|^2 (1-|\xi_1|^2) \Bigr ).
\label{Scoh}
\eeq

\subsection{Delocalized-boson ground state}
For $U=0$ ($\tau \to \infty$), the ground state is found by setting $\xi_1= \xi_2= \xi_3= 1/\sqrt 3$ in formula (\ref{state}).
This gives simply
\beq
|{\rm del }\rangle=
\frac{1}{\sqrt{3^N\,N!}}( \hat{a}_1^\dagger+\hat{a}_2^\dagger+\hat{a}_3^\dagger)^N|0\rangle
\label{0}
\;,\eeq
which is state (\ref{coherentclosed}) describing the boson delocalization.
From formula (\ref{Scoh}), with $|\xi_1|^2= 1/3$ and large $N$, we find that the entanglement entropy in the
delocalized-boson regime is given by
\beq
S_{del}\simeq \frac{1}{2}\log_{2}\left(4\pi e N\right)-\log_{2} 3.
\label{S0}
\eeq
This value exactly reproduces the asymptotic value obtained numerically
for $\it S$ when $\tau \gg 1$.

\subsection{Cat-like ground state}

The strong-interaction regime ($|U| \gg J/N$) is characterized by a ground state representing
a cat-like state. The latter, tending to the NNN state (\ref{w}) for $\tau \to 0 $,
can be constructed by exploiting state (\ref{state}).
Classically, for $\tau$ small enough, the self-trapping effect
entails an almost complete localization of attractive bosons at some site (soliton-like state).
Quantum-mechanically, the peak of the boson population can be placed, with the same probability,
in one of the three sites of ring.
A realistic ansatz for this state thus consists in superposing
three states (\ref{state}) each one involving the localization at a different site. One has
\begin{eqnarray}
|{\rm cat} \rangle = \frac{1}{\sqrt{3\,N!}}
\sum^3_{k=1}
\bigl ( \xi {\hat a}_k^\dagger+ \eta ( {\hat a}_{k+1}^\dagger
+ {\hat a}_{k+2}^\dagger) \bigr )^N |0\rangle
\label{cat}
\; ,
\end{eqnarray}
where ${\hat a}_{k+3} = {\hat a}_k$ within the periodic ring geometry. The $k$th state
$
|\Psi_k \rangle =
\bigl ( \xi {\hat a}_k^\dagger + \eta ( {\hat a}_{k+1}^\dagger+ {\hat a}_{k+2}^\dagger )
\bigr )^N |0\rangle/\sqrt {N!}$
of this superposition features local populations
$$
\langle \Psi_k | {\hat n}_k |\Psi_k \rangle = N |\xi|^2
\, , \quad
\langle \Psi_k | {\hat n}_i |\Psi_k \rangle = N |\eta|^2\, ,
$$
for $i = k+1, k+2$.
In each $|\Psi_k \rangle$, well $k$ exhibits a population larger than the those of wells $k+1$ and $k+2$,
namely $|\xi_k|^2 \equiv |\xi|^2 > |\xi_{k+1}|^2 = |\xi_{k+2}|^2 \equiv |\eta|^2$, where
$\xi_j$ are the coherent-state parameters of definition (\ref{state}).

For $N \gg 1$ the coherent-state product reduces to the simple form
$\langle \Psi_h |\Psi_k \rangle \simeq \delta_{hk}$ while in the energy expectation value
$$
E = \langle {\rm cat} | {\hat H} | {\rm cat} \rangle =
\frac{1}{3} \sum_k\sum_h \langle \Psi_k | {\hat H} |\Psi_h \rangle
$$
mixed terms $\langle \Psi_k | {\hat H} |\Psi_h \rangle $ with $h \ne k$ can be shown \cite{vittorio1}
to vanish for large $N$.
In addition, due to the translation simmetry of $\hat H$,
$\langle \Psi_k | {\hat H} |\Psi_k \rangle$ yields the same expression for each $k$. Then,
for $k=1$, one finds $E = \langle {\rm cat} | {\hat H} | {\rm cat} \rangle$
$\simeq \langle \Psi_1 | {\hat H} |\Psi_1 \rangle$
giving
$$
E \simeq
\frac{U}{2} N(N-1) (|\xi|^4 +2 |\eta|^4) -2J N ( \xi \eta^* + \xi^* \eta + |\eta|^2)\, .
$$
Interestingly,
the latter coincides with the energy obtained from formula (\ref{scenergy})
when setting $\xi_1 = \xi$ and $\xi_3 =\xi_2 =\eta$.
This makes it evident that the effective energy $E$ is essentially independent from the symmetrized
form of cat-like state (\ref{cat}). By defining $z=|\xi|^2-|\eta|^2$ and the inverse formulas
\beq
|\xi|^2 = \frac{2z +1}{3} \, ,\quad |\eta|^2 = \frac{1-z}{3}\, ,
\label{xieta}
\eeq
one obtains
$$E \simeq\frac{U}{6}\,N\,(N-1)\Bigl [1+2\,z^2 \Bigr]\;
\qquad\qquad\qquad\qquad
$$
\vspace{-0.5cm}
\beq
- \frac{2\,J}{3}N \Bigl (1-z + 2 \sqrt{(1-z)(1+2\,z)} \, \cos \phi_{12}  \Bigr )
\; ,
\label{efinal}
\eeq
where $\phi_{12} = \phi_{1}- \phi_{2}$ is defined by formula (\ref{phase}). Note that
only such a relative phase survives in $E$ since $\xi_2 = \xi_3 \equiv \eta$ in $|\Psi_1 \rangle$.
In order to identify the energy minimum one must impose $\phi_{12}=0$, phase $\phi_{12}$ being
independent from $z$. The stationary values of energy (\ref{efinal})
in the attractive case $U<0$, for $N\gg 1$, are found by 
the solutions of
\beq
\label{equa}
\frac{dE}{dz}\simeq  -\frac{2}{3}|U|N^2 \left [
z-\tau - \frac{\tau( 4z-1)}{ {\sqrt {1-z} } {\sqrt {1+2z} }} \right ] =0
\, ,
\eeq
with $\tau = {J}/{(|U|N)}$. For $\tau \ll 1$, in addition to $z=0$, equation (\ref{equa}) exhibits two
solutions one of which is negative. In this regime,
these are necessarily placed in the proximity of asymptotes $z=1$ and $z=-1/2$ of function $dE/dz$.
The negative solution $z_1 \simeq -1/2$ can be discarded in that, owing to $|\xi|^2 \ll |\eta|^2$,
it corresponds to an antisoliton state (this typically represents an excited state)
where one of the three wells
is essentially depleted. In this regime solution, $z=0$ (which satisfies equation (\ref{equa})
for any $\tau$) and $z_1$ correspond to an energy maximum and to a local minimum, respectively.
This can be easily checked through the calculation of the second derivative ${\ddot E } = d^2E/dz^2$.
The positive solution can be found by substituting $z = 1 - v$ (with $v \ll 1$) in equation (\ref{equa}).
This gives
$$
z= z_2 \simeq 1 - \frac{3\, \tau^2}{(1-\tau)^2}\,\, \to \,\,\,
|\xi|^2 \simeq  1-2\tau^2\, ,\,\, |\eta|^2 \simeq \tau^2\, ,
$$
which can be shown to represent the effective minimum of $E (z)$ being ${\ddot E } > 0$.
The fact that the energy minimum entails $|\xi|^2 \gg |\eta|^2$ is consistent with
the distribution of bosons prescribed by cat-like state (\ref{cat}). This validates
the ground-state representation based on formula (\ref{cat}).

The entanglement entropy is found by assuming that the reduced density matrix,
in the large $N$ limit, is formed by a symmetric linear combination of three almost-normal
distributions, i.e.
\beq
\hat\rho_{cat} = \frac{1}{3}\bigl ( \hat\rho_{1} (\xi_1)+ \hat\rho_{1}(\xi_2)+\hat\rho_{1}(\xi_3) \bigl)
\label{rhocat}
\eeq
(matrix $\hat\rho_{1}$ is defined by formula (\ref{densitymatrix}))
reflecting the parameter choice $\xi_1=\xi$, $\xi_2=\xi_3=\eta$ of coherent state $|\Psi_1 \rangle$.
In the extreme case of the NNN state, namely for $z=1$, the entropy is equal to $(\log_2 3-2/3)$.
Interestingly, the same result is achieved by calculating the single-site entropy
$S = - \sum^N_{n=0} \rho_{1,n} {\log}_2  \rho_{1,n}$ for the state NNN where
$\rho_{1,0} = 2/3$, $\rho_{1,N} = 1/3$ and $\rho_{1,n} = 0$ otherwise. This seems to confirm the validity
of the form assumed for the density matrix $\hat\rho_{cat}$.
More in general, for a generic cat-state in the large $N$ limit, we can write
\beq
\label{Scato}
S_{cat}\simeq \log_2 3-{2}/{3} + S(\xi)/3+2S(\eta)/3
\;,
\eeq
where $S(\alpha)=- Tr_1 \left(\hat\rho_{1}(\alpha)\,{\log}_2\, \hat\rho_{1}(\alpha)\right)$, with $\alpha = \xi, \eta$,
are the single-site entropies for sites populated by $N|\xi|^2$ and $N|\eta|^2$ bosons, respectively.
In the limit $z \to 1$ these entropies are zero since they corresponds to Fock states. Hence we recover the result for the NNN
state. Instead, as soon as $1-z\gg 1/N$ (where $z$, however, must ensure a
population $|\xi|^2$ considerably larger than $|\eta|^2$),
$S(\xi)$ and $S(\eta)$ are given by Eq. (\ref{Scoh}) with $\xi_1=\xi$ and $\xi_1=\eta$. In this case Eq. (\ref{Scato}) becomes
\beqa
\nonumber S_{cat}&\simeq& \frac{1}{6}\log_2\Bigl [ 4\pi e N(1-z)(1+2z) \Bigr ] \\
&+&\frac{1}{3}\log_2\Bigl [ 2\pi e N(1-z)(2+z) \Bigr ]-\frac{2}{3}.
\label{Scat}
\eeqa

\subsection{Semiclassical interpretation of the transition to the cat-like state }
\label{subVB}

Equation (\ref{equa}), derived from energy (\ref{efinal}),
supplies further interesting
information about the transition from the cat-like to the uniform state for $\tau \to 1/4$.
If $\tau$ is increased, the negative solution $z_1$ approaches solution $z=0$ and
for $\tau = 2/9$ the two solutions coincide. This value is particularly significant
because solutions $z_1$ and $z=0$ change their character when $z_1$ crosses $z=0$.

For $\tau > 2/9$ solution $z_1$ becomes positive while $z_2 < 1$ moves away from $1$.
By assuming that $z_1$ is positive but small,  equation (\ref{equa}) supplies
the approximate expression $z_1 = 3(\tau -2/9)/\tau$ giving ${\ddot E } (z_1) <0$.
Then $z_1 >0$ corresponds to an energy maximum.
In parallel, a second minimum crops up at $z=0$,
in addition to the energy minimum associated to solution $z_2$.
At $z=0$ the second derivative is found to be
${\ddot E } (0) = -2|U| N^2 (1-9\tau /2)/3$ confirming that this solution
becomes a minimum if $\tau > 2/9$.
The relevant populations, given by formulas (\ref{xieta}) with $z=0$,
show how the second minimum involves the uniform distribution
corresponding to state (\ref{coherentclosed}).

The occurrence of a second energy minimum when parameter $\tau$ is varied
is extremely interesting.
It is the signal that, within the present coherent-state picture of $|\Psi \rangle$,
the system manifests a second way to realize a minimum-energy configuration
in the range $2/9 \le \tau \le 1/4$. In this interval
the numerical values of $E(0)$ and $E(z_2)$ are very close. For $\tau = 1/4$
the exact solutions of equation (\ref{equa}) can be easily found. These are
$z_1 = 1/4$ and $z_2 = 1/2$ in addition to $z=0$. In this special case
$E(0)=E(z_2)$.

This behavior, emerging in the interval $[2/9,1/4]$, can be viewed as the
semiclassical counterpart of the considerable changes characterizing the ground-state structure
shown in Fig.~\ref{fig1} for $0.2020 \le \tau \le 0.2415$.
The presence of two (semiclassical) energy minima corresponds, quantum-mechanically,
to the coexistence in the ground state of two dominating components noted
in Ref. \cite{vittorio1}: the cat-like state
(with three peaks describing the three possible ways to get a complete localizations) and
state (\ref{coherentclosed}) characterized by the boson delocalization.

\subsection{Coexistence of $|{\rm cat} \rangle$ and $|{\rm del} \rangle$}

The previous discussion in subsection \ref{subVB} and the results illustrated in Fig.~\ref{fig1}
suggest that, in the regime classically identified by $2/9 \le \tau \le 1/4$, the ground-state
features the coexistence of a cat state and a (delocalized) coherent state.
Explicitly, our ansatz for the ground state $|\Psi\rangle$ is
\beq
|{\rm mix} \rangle = \sqrt{1-\alpha^2} \, |{\rm cat} \rangle +  \alpha\, | {\rm del } \rangle
\label{X}
\; ,
\eeq
where state $|{\rm del }\rangle$ is given by (\ref{coherentclosed}) and
$\alpha = 1/{\sqrt{2}}$ has been assumed. Moreover, we consider the state $|{\rm cat}\rangle$
corresponding to solution $z=z_2= 1/2$ of case $\tau = 1/4$, where
parameters $\xi$ and $\eta$ are such that $|\xi|^2 = 2/3$ and $|\eta|^2 = 1/6$.
By construction the overlap between state $|{\rm cat} \rangle$ and
the localized coherent state $|{\rm del} \rangle$
is small and become negligible in the limit of large $N$. This follows when
the formula describing the product of two coherent states
$
\langle \xi | \zeta \rangle = \bigl ({\sum}_{i=1}^3 \xi^*_i \zeta_i \bigr)^N
$
is applied to the product of $|{\rm del} \rangle$ with one of the three components of $|{\rm cat} \rangle$,
and among the three components of $|{\rm cat} \rangle$ themselves.
Under this assumption, the entanglement entropy is given by
\beqa
\nonumber
S_{mix}
&\simeq& (1-\alpha^2)\, S_{cat}{(z=1/2)}+\alpha^2 S_{del}\\
&&-(1-\alpha^2) \log_2(1-\alpha^2)-\alpha^2\log_2\,\alpha^2
\label{SXalpha}
\;,\eeqa
where $S_{cat}(z=1/2)$ is given by Eq. (\ref{Scat}) by setting
$z = 1/2$ and $S_{del}$ is given by Eq. (\ref{S0}). For $\alpha^2=1/{2}$, Eq. (\ref{SXalpha}) becomes simply
\beq
\label{SX}
S_{mix}\simeq
\frac{1}{2}\log_{2}\big({4}\pi e N/3\big)+ \frac{1}{6}\big(1+\log_{2} 5\big) \,.
\eeq
Notice that $S_{mix}$ is always greater than the entropy of the cat state, Eq.(\ref{Scat}),
for any value of $z$. For $N=45$, $S_{mix} \simeq 5.05$,
which is in a quantitatively good agreement with the maximum value achieved by $S$,
found through the exact diagonalization of the BH Hamiltonian
(see the comments about Fig.~\ref{fig2}).


\section{Conclusions}
We have studied a system of interacting dipolar bosons confined by a triple-well potential
with periodic boundary conditions. The latter can be described in terms of a 3-site extended
Bose-Hubbard (BH) model including the boson interaction $U_1$ between nearest-neighbor sites.
Thanks to its symmetry properties, we have reduced the dipolar-boson model to the symmetric BH
model depending on a unique (effective) interaction $U$.
Experimentally, this fact is certainly interesting because one can explore the attractive
regime ($U<0$) of triple-well dynamics by using positive-valued onsite and nearest-neighbor
interactions ($U_0$ and $U_1$, respectively). This circumstance allows one to study the
attractive regime of BH model avoiding the
possible collapse due to $U_0 <0$ for sufficiently large atomic densities.

We have analyzed the ground-state structure of dipolar-bosons in a triple-well
by changing the (unique) parameter $U$ in its whole variation range.
Fig.~\ref{fig1} illustrates the numerical study of the ground-state structure for different $U$.
In the range $U>0$, the ground state has been shown to exhibit the expected Mott-insulator (superfluid)
form for $\tau=J/(N|U|) \ll 1$ ($\tau \gg 1$).
For attractive interaction $U<0$, the ground state goes from
a su(3) coherent state (describing the boson delocalization of the superfluid state)
to a macroscopic superposition of three states with localized populations (the cat-like state)
separated by an intermediate range where the ground state has a mixed character.
The latter is well described by a linear combination of the delocalized-boson ground state and
the macroscopic cat-like state.
The semiclassical study of the low-energy landscape in the cat-like regime shows that the energy
features the coexistence of two minima which can be viewed as the classical counterpart of the ground
state with a mixed character.

To stress the importance of the quantum correlations between different sites,
we have investigated the single-site entanglement entropy as a function of the
effective interaction. This analysis has been carried out both numerically, by
diagonalizing the extended BH Hamiltonian, and analytically, by representing
the ground state in terms of coherent states.
In particular, we have found that the single-site entanglement entropy $S$ reaches its maximum
within the coexistence regime. The $S$ maximum is attained in correspondence to the transition
from the four-peak to the three-peak structure of the ground state. In parallel, the $\tau$ derivative
$S$ is found to have a maximum and a minimum point in the vicinity of which $S$ undergoes a rapid
change. The rapid increasing of $S$ takes place in the regime heralding the NNN state, while the
rapid decreasing of $S$ corresponds to the merging of the lateral peaks with the central one.

\acknowledgments


This work has been supported by MIUR (PRIN 2010LLKJBX). 
LD, GM and LS acknowledge financial support from 
the University of Padova (Progetto di Ateneo 2011) and  
Cariparo Foundation (Progetto di Eccellenza 2011). 
LD acknowledges financial support also from MIUR (FIRB 2012 RBFR12NLNA).


\end{document}